\documentclass{ws-procs9x6}

\usepackage{citesort,epsfig}

\begin{document}

\title{Photo-excitation of hyperons and exotic baryon in
$\gamma N \to K \overline{K} N$
\footnote{\uppercase{W}ork supported by
\uppercase{F}orschungszentrum-\uppercase{J}{\"u}lich, contract \uppercase{N}o.
41445282 (\uppercase{COSY}-058) and
\uppercase{U.S. DOE N}uclear \uppercase{P}hysics \uppercase{D}ivision
\uppercase{C}ontract \uppercase{N}o. \uppercase{W-31-109-ENG-38}.}}

\author{Yongseok OH and K. NAKAYAMA}

\address{Department of Physics, University of Georgia, Athens, Georgia 30602,
U.S.A.}

\author{T.-S.H. Lee}

\address{Physics Division, Argonne National Laboratory, Argonne,
Illinois 60439, U.S.A.}

\maketitle

\abstracts{
We investigate the reaction of $\gamma N \to K \overline{K} N$ focusing
on the photoproduction of $\Lambda(1520)$ and putative exotic
$\Theta(1540)$.
We consider various background production mechanisms including the production
of vector mesons, tensor mesons, and other $\Lambda$ and $\Sigma$ hyperons.
We discuss the angular distribution of $\Lambda(1520)$
photoproduction cross section and the radiative decays of $\Lambda(1520)$.
We also discuss what we expect for the invariant mass
distributions if the $\Theta(1540)$ is formed in the reaction,
with the parameters studied so far.
We find that the peak in the $KN$ invariant mass distribution, if confirmed, can
hardly come from the kinematic reflections, especially, due to the tensor meson
backgrounds.
}

Recent experimental activities searching for exotic pentaquark
states\cite{LEPS03,CLAS03-b} renewed the interests in
double kaon photoproduction, i.e., $\gamma N \to K \overline{K} N$,
motivating recent theoretical investigations on its production
mechanisms\cite{NT03,DKST03,Roberts04,ONL04}.
Investigating various physical quantities of this reaction allows the
study of hadron resonances that are formed during the reaction.
For example, from its $K\overline{K}$ channel, we can
study the photoproduction of mesons, which decay mostly into $K\overline{K}$,
and the $\overline{K}N$ channel is directly related to the $S=-1$
hyperon production, where $S$ is the strangeness.
Furthermore, the $KN$ channel of this reaction gives a tool of searching
for the hypothetical $S=+1$ exotic baryons.
In fact, this is the reaction used by LEPS Collaboration\cite{LEPS03}
to observe the exotic $\Theta$, which caused a lot of theoretical and
experimental investigations\cite{ONL04}.
However, the signals for $\Theta(1540)$ could not be confirmed by other
experiments, which lead to a doubt on its existence.

Another motivation of this work is the radiative decays of hyperon
resonances.
As stated above, double kaon photoproduction is used to study the
production of hyperon resonances.
Close inspection of the kinematic region of this reaction at low energies
shows that the $\phi$ meson production is dominant in the $K\overline{K}$
channel and $\Lambda(1520)$ is in the $\overline{K}N$ channel.
Therefore, this reaction may be used for studying the production
mechanisms of $\Lambda(1520)$, which contain the radiative decays of
$\Lambda(1520)$.
Theoretical predictions on the radiative decays of $\Lambda(1520)$ are
strongly model-dependent and their experimental data are very limited and
uncertain\cite{PDG04}.
There are recent measurements for $\Gamma(\Lambda(1520) \to
\Lambda\gamma)$ by SPHINX Collaboration and CLAS
Collaboration\cite{SPHINX04a,CLAS05a}, of which results are consistent to
each other.
But the decay width of $\Lambda(1520) \to \Sigma\gamma$ has not been
reported so far.
(Note that the decay width of $\Lambda(1520) \to \Sigma\gamma$
in PDG\cite{PDG04} is not from a direct measurement, but
from that of $\Lambda(1520) \to \Lambda\gamma$ by using some
SU(3) relations.)

In this work, we consider the photoproduction of $\Lambda(1520)$ and
that of $\Theta(1540)$.
As background production mechanisms of the latter, we consider vector meson
production,
tensor meson production, and the $t$-channel (tree) Drell diagrams, as well as
the production of other $\Sigma$ and $\Lambda$ hyperons.
We also include the form factors and constrain the cutoff parameter by
the total cross section data for $\gamma p \to K^+ K^- p$.
The details on the diagrams, effective Lagrangians used in this
calculation, and the method to restore the charge conservation condition
can be found in Ref.~\refcite{ONL04}.
The models for vector meson production can be found in Ref.~\refcite{OTL}.

\begin{figure}[t]
\centering
\epsfig{file=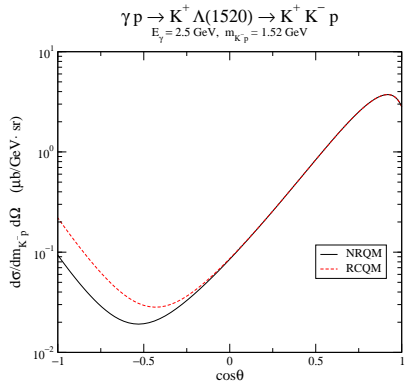, width=0.5\hsize,angle=-90}
\caption{Differential cross section for $\gamma p \to K^+ \Lambda(1520)
\to K^+ K^- p$. The solid and dashed lines are obtained by using the
nonrelativistic quark model\protect\cite{KMS85} and relativistic quark
model\protect\cite{WPR91} predictions for the
$\Lambda(1520) \to \Sigma\gamma$ decay.}
\label{fig1}
\end{figure}

Shown in Fig.~\ref{fig1} are the results for the double differential cross
sections for $\gamma p \to K^+ \Lambda(1520) \to K^+ K^- p$, where
$\theta$ is the angle of $K^+$ in the c.m. frame.
In this calculation we use the value of CLAS Collaboration for the
$\Lambda(1520) \to \Lambda\gamma$ decay\cite{CLAS05a}, $\approx 167$ keV.
For the decay of $\Lambda(1520)$ into $\Sigma\gamma$ we take two model
predictions:
the nonrelativistic quark model prediction of Ref.~\refcite{KMS85},
$\approx 55$ keV,
and that of the relativistic quark model of Ref.~\refcite{WPR91},
$\approx 293$ keV.
The result shows that the double differential cross section at large
scattering angles depends on the decay width of $\Lambda(1520) \to
\Sigma\gamma$.

\begin{figure}[t]
\centering
\epsfig{file=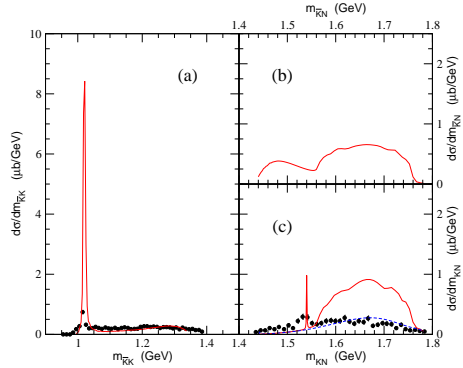, width=0.5\hsize, angle=-90}
\caption{
(a) $K\overline{K}$, (b) $\overline{K}N$, and (c) $KN$ invariant
mass distributions for $\gamma n \to K^+ K^- n$ at $E_\gamma = 2.3$ GeV.
The data are from Ref.~\protect\refcite{CLAS03-b}.
The dashed line in (c) is obtained without the $\phi$ and $\Theta$ 
contributions.}
\label{fig2}
\end{figure}

Next, we consider the production of exotic $\Theta(1540)$.
For this calculation, we assume that the $\Theta$ belongs to
$\overline{\bf 10}$, and it has $J^P = \frac12^+$ and a decay width of
1~MeV\cite{PDG04}.
The SU(3) effective Lagrangian for the interaction of $\overline{\bf
10}$ with the normal baryon and meson octet can be found, e.g., in
Ref.~\refcite{LKO04-OK04}.
In Fig.~\ref{fig2}, we present the results for the invariant mass
distributions of various channels in the $\gamma n \to K^+ K^- n$ reaction.
Apparently, the peak coming from the $\Theta(1540)$ formation depends on
the cross section of $\Theta$ photoproduction.
In this work, we used the model of Ref.~\refcite{OKL03a} for
$\gamma N \to \overline{K} \Theta$.
(See, e.g., Ref.~\refcite{NL04} for the general properties of this reaction.)

One interesting question about this reaction is whether the peak in the $KN$
channel could come from the other backgrounds, especially from tensor meson
production\cite{DKST03}.
This was motivated by an old experiment\cite{ABBC69} on
$\pi^- p \to K^- X$, where the observed peaks in the exotic channel were
ascribed to the background, especially higher-spin meson, production mechanisms,
since the peak positions moved by changing the beam energy.
Therefore, it is crucial to test whether this can explain the peak in the
$KN$ channel of double kaon photoproduction.
In this work, we include $a_2^0(1320)$ and $f_2(1275)$ tensor meson
production as backgrounds.
We found that (i) the tensor meson production alone gives a very broad
peak not a sharp peak, although we confirmed that the peak position moved
depending on the beam energy, and (ii) the tensor meson production part is
suppressed very much compared with the other reaction mechanisms, and its
contribution is hard to be seen even after removing the $\phi$ and
$\Lambda(1520)$ production part from the backgrounds.
Therefore, any sharp peak, if confirmed, can be hardly explained by kinematic
reflections.
In addition, a very sharp peak is expected in the $KN$ invariant
mass distribution if the exotic $\Theta$ is formed in the reaction.
(We note that the most recent CLAS experiment\cite{CLAS05b} shows no
evidence of $\Theta(1540)$, but its existence is still 
controversial\cite{STAR05,Burkert05}.)

\end{document}